\documentclass[conference]{IEEEtran}
\usepackage{cite}
\usepackage{textcomp}
\usepackage[pdftex]{graphicx}
\DeclareGraphicsExtensions{.pdf,.jpeg,.png}
\usepackage{amsmath}
\usepackage{algorithm,algorithmic}
\usepackage{xcolor,soul}
\usepackage{url}
\usepackage{fancyhdr}
\usepackage{textcomp}
\fancypagestyle{empty}{%
  \fancyhf{}% Clear header/footer
  \fancyhead[C]{Accepted as conference paper at the ISCAS 2020}
  \fancyfoot[C]{\footnotesize{This document is the paper as accepted for presentation at the IEEE International Symposium on Circuits and Systems (ISCAS) 2020.\\
\textcopyright 2020 IEEE. Personal use of this material is permitted. Permission from IEEE must be obtained for all other uses, in any current or future media, including reprinting/republishing this material for advertising or promotional purposes, creating new collective works, for resale or redistribution to servers or lists, or reuse of any copyrighted component of this work in other works.}}% Your journal/note
}

\begin{document}
%\title{Experimental Demonstration of Binarized ADALINE network on OxRAM Crossbar}
\title{Methodology for Realizing VMM with Binary RRAM Arrays: Experimental Demonstration of Binarized-ADALINE Using OxRAM Crossbar\\[-1.2ex]}
%\title{Experimental Demonstration of Binarized Classifiers on RRAM Crossbar: Binarized ADALINE Implementation}
\author{\IEEEauthorblockN{Sandeep Kaur Kingra$^1$,Vivek Parmar$^1$,Shubham Negi$^1$,Sufyan Khan$^1$,Boris Hudec$^2$,Tuo-Hung Hou$^2$ and Manan Suri$^{1}$}
\IEEEauthorblockA{$^1$Indian Institute of Technology Delhi, Hauz Khas, New Delhi - 110016, India, Email ID : manansuri@ee.iitd.ac.in\\
%$^2$ CYRAN AI Solutions Pvt. Ltd. , New Delhi - 110016, India\\ 
$^2$ National Chiao Tung University,Hsinchu 300, Taiwan\\ 
}\\[-4ex]}

\maketitle
\begin{abstract}
In this paper, we present an efficient hardware mapping methodology for realizing vector matrix multiplication (VMM) on resistive memory (RRAM) arrays. Using the proposed VMM computation technique, we experimentally demonstrate a binarized-ADALINE (Adaptive Linear) classifier on an OxRAM crossbar. An 8$\times$8 OxRAM crossbar with Ni/3-nm HfO$_2$/7 nm Al-doped-TiO$_2$/TiN device stack is used. Weight training for the binarized-ADALINE classifier is performed ex-situ on UCI cancer dataset. Post weight generation the OxRAM array is carefully programmed to binary weight-states using the proposed weight mapping technique on a custom-built testbench. Our VMM powered binarized-ADALINE  network achieves a classification accuracy of 78\% in simulation and 67\% in experiments. Experimental accuracy was found to drop mainly due to crossbar inherent sneak-path issues and RRAM device programming variability. 
\end{abstract}

% \fi
\IEEEpeerreviewmaketitle
\thispagestyle{empty}
\section{Introduction}
In-Memory Computing (IMC) and analog hardware Vector Matrix Multiplication (VMM) approaches offer efficient alternatives to conventional computing for resource hungry modern-AI workloads \cite{Shafiee_2016}. Advance neural networks such as Convolutional Neural Networks (CNNs) and Recurrent Neural Networks (RNNs) involves extensive use of VMM operations. In particular, emerging resistive memory (RRAM) nanodevice based arrays offer an efficient and compact option for performing VMM operations in hardware. Several research groups \cite{Hu_2016,Truong_2016,Lastras_Montano_2017,Jeong_2018,Sun_2019} have demonstrated this analog computing method for a variety of applications such as linear equation solver \cite{Sun_2019}, image processing \cite{Hu_2011}, data compression \cite{Yuhao_Wang_2015}, feature extraction \cite{Choi_2017}, neural network inference \cite{Alibart_2013}, in-situ training \cite{Alibart_2013,Mondal_2018}, etc. Multiple emerging memory nanodevice technologies have been utilized for this application : OxRAM \cite{Shafiee_2016}, MRAM \cite{Mondal_2018}, PCM \cite{Wang_2019}, Ferroelectric FET\cite{Jerry_2017}, ECRAM \cite{Tang_2018}, Flash \cite{Mahmoodi_2018}, etc. Owing to their higher device density, crossbar structures are preferred for VMM applications \cite{Tan_2018}. Typical implementations of crossbar based VMM operations utilize analog conductance states of RRAM nanodevices \cite{Shafiee_2016,Hu_2016,Lastras_Montano_2017,Tsai_2018}. However, difficulty to obtain multiple reliably programmable resistance states, non-linearity of analog RRAM device conductance and variability related issues pose a major challenge to this approach. Furthermore complex program-and-verify schemes may be required in certain cases\cite{Moon_2019}. In case of selector-free crossbars these issues are further enhanced due to the presence of sneak-paths\cite{Cassuto_2016}. Binary neural networks (BNNs) have been shown to have better performance (i.e. lower memory, time and energy requirements) \cite{courbariaux2016binarized,Liang_2018}, compared to their full-precision (analog) counterparts, at the cost of a marginal accuracy trade-off. Hence using crossbar based VMM for hardware realization of BNN would face lesser challenges compared to fully analog alternatives. Furthermore, use of binary memory states leads to simplification of programming with relatively less impact on device endurance. Recently BNNs have been demonstrated on RRAM matrix using 2T-2R synaptic cells \cite{Huang_2019}. However, true density benefit of emerging RRAM technology can be exploited only by using selector-free RRAM crossbars. In the current study, we propose a hardware mapping methodology for realizing VMM using binary RRAM crossbars. Further we experimentally demonstrate the proposed technique with a case-study of binarized-ADALINE using OxRAM crossbar. To the best of our knowledge, this is the first experimental validation of BNNs using selector-free RRAM crossbars.

Compared to literature we present the following novel concepts in this paper:
\begin{enumerate}
    \item Weight mapping methodology, algorithm and operation scheduling for implementing BNN on any RRAM array (ex- OxRAM, CBRAM, PCM, etc.).
    \item Experimental demonstration of a binarized-ADALINE network on an 8$\times$8 OxRAM crossbar for classification on UCI cancer dataset \cite{Dua_2019}.
\end{enumerate}
\begin{figure}[t]
\centering
  \includegraphics[width=\linewidth]{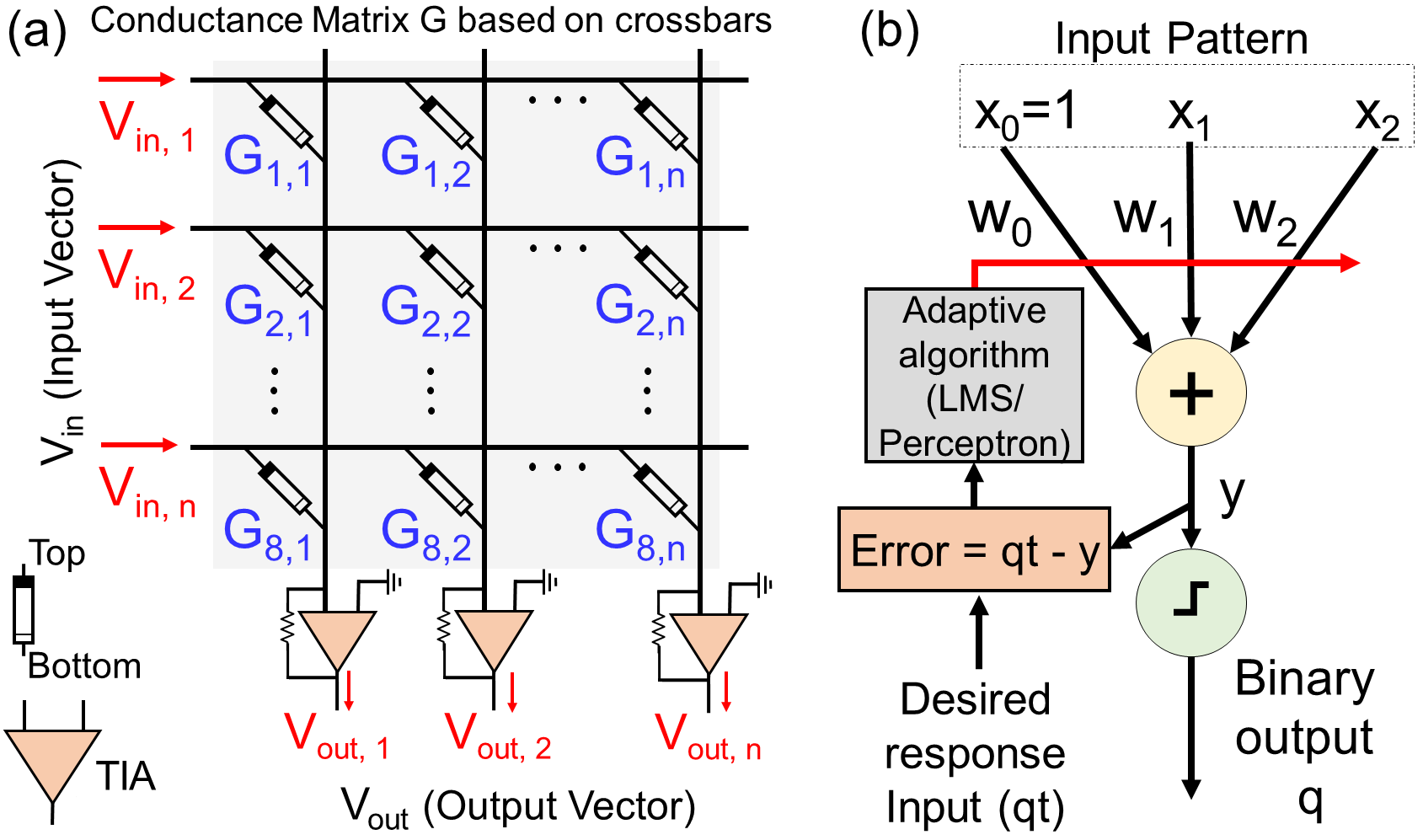}
  \caption{(a) Basic VMM operation implemented using two-terminal resistive nanodevice crossbar. (b) Structure of generic ADALINE Network.}
  \label{Fig1a}
\end{figure}
The paper is organized as follows: Section \ref{sec2} describes basic concepts relevant for this work. Section \ref{sec3} describes the proposed methodology for mapping BNNs and training a binarized-ADALINE network. In Section \ref{sec4}, we describe the custom-test platform in detail along with experimental results on fabricated OxRAM crossbar.

\section{Basics and Background}
\label{sec2}
\subsection{Vector Matrix Multiplication (VMM) in Hardware}
Fig. \ref{Fig1a}(a) describes the standard implementation of VMM operation in a generic two-terminal resistive nanodevice crossbar based architecture. Input is applied in the form of voltages ($V_{in,i}$) across all rows in parallel. Based on the conductance state ($G_{i,j}$) of the devices current integrates on each column. The resulting integrated current is converted to voltage using a Trans-Impedance Amplifier (TIA) with feedback resistor ($R_f$). %Here we assume the device behaves in an ideal fashion and no sneak path based noise influences the rows. 
Output voltage ($V_{out,j}$) or the output of VMM operation (across a given column j of the matrix) is defined by Eq. \ref{eq1a}: 
\begin{equation}
\label{eq1a}
    V_{out,j} = R_f \times V_{in,i} \times G_{i,j} 
\end{equation}

\subsection{ADALINE}
`Adaptive Linear Element' also known as ADALINE, was one of the first networks proposed based on "memistors" (not memristors) \cite{Widrow_1990}.  It is used for a variety of classification applications such as power-quality event detection \cite{abdel2003power}, stereo-vision matching \cite{Pajares_2001}, etc. that require fast computation. It is an adaptive threshold logic element representative of a neuron used in ANNs (Artificial Neural Networks). As shown in Fig. \ref{Fig1a}(b) it consists of an adaptive linear combiner, cascaded with a hard-limiting quantizer, which is used to produce a binary output, $Y_k$ = sign ($s_k$). The bias weight $w_{0k}$ which is connected to a constant input $x_0$ = +1 effectively controls the threshold level of the quantizer. In single-element neural networks, an adaptive algorithm such as Least Mean Square (LMS) or Perceptron rule is used for training weights of ADALINE. 

\section{Proposed Methodology for VMM-based BNN}
\label{sec3}
\begin{figure}[t]
  \centering
  \includegraphics[width=\linewidth]{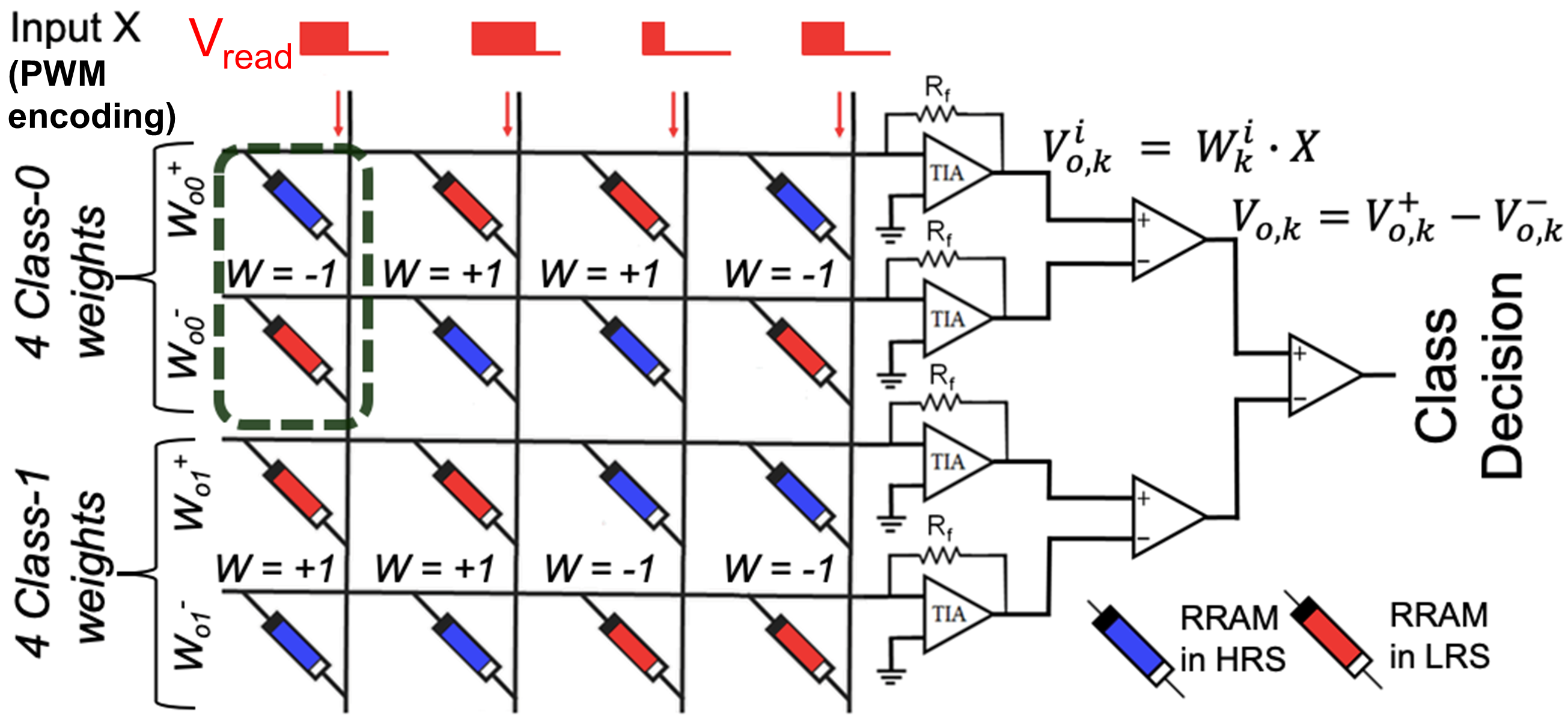}
  \caption{Proposed scheme for computation and weight mapping using two-terminal resistive crossbar. Every logical weight value (i.e. +1 or -1) is mapped on the crossbar using 2 paired devices from consecutive rows (i.e. rows W$^+$ and W$^-$) of the same column. The paired devices are always programmed to complementary states (LRS-HRS or vice-versa). In particular, to realize logical weight +1, device from the first row is programmed to LRS and the paired device in the consecutive row is programmed to HRS. For realizing logical weight -1, programming is inverted (i.e. first row device is in HRS while consecutive row device is in LRS). Eight logical weight values (-1,1,1,-1,1,1,-1,-1) are programmed using 16 (4x4) devices. The first four weights corresponding to class k=0 (-1,1,1,-1) are partitioned in rows 1 and 2, while the next four weights corresponding to class k=1 (1,1,-1,-1) are partitioned in rows 3 and 4. Note, input voltages are applied on columns and current integration occurs across rows. This is due to the fact that DAC units used in our experimental setup generate only positive voltages. Negative voltages are effectively realized by grounding device top-electrode and applying +ve DAC signal at device bottom electrode. Programming and read paths are isolated using CMOS switches for each channel.}
  \label{Fig1}
\end{figure}

\begin{figure*}[!ht]
  \centering
  \includegraphics[width=0.9\linewidth]{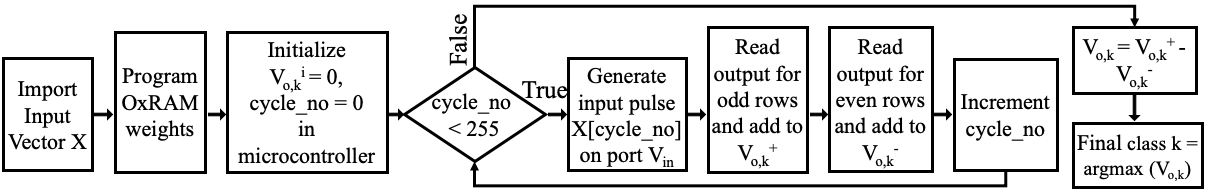}
  \caption{Flowchart summarizing sequence of operations to perform binarized-ADALINE computation.}
  \label{Fig1b}
\end{figure*}
\subsection{Training for Binarized-ADALINE}
We use a modified version of the training algorithm proposed in \cite{Moons_2017}. %by incorporating support for ADALINE \textcolor{blue}{and generic datasets beyond images. 
The specific modifications include: (i) Training with only positive value inputs. This helps to simplify input encoding scheme compared to inputs with signed magnitudes as normally used for training BNNs \cite{courbariaux2016binarized}. (ii) Adding support for training a binarized-ADALINE network by using binary `tanh' as the hard-limiting quantizer and binary weights. (iii) Support for training non-vision datasets by use of min-max normalization on input features.
%We have also enabled support for training with only positive value inputs in order to simplify input encoding scheme.} 
The training algorithm is summarized in Algorithm \ref{algo1}. Post-training, binary weights (+1,-1) are generated.

\begin{algorithm}[b]
 \algsetup{linenosize=\tiny}
 \scriptsize
 \caption{Algorithm for training binarized-ADALINE}
 \begin{algorithmic}[1]
 \renewcommand{\algorithmicrequire}{\textbf{Input:} \begin{minipage}[t]{0.8\linewidth}{Training data X, Expected target Y, Epochs N, Batch-size B\vspace{0.5em}}\end{minipage}}
 \renewcommand{\algorithmicensure}{\textbf{Output:} Target Y$_t$}
 \REQUIRE 
 \ENSURE  .
 \\ \textit{Initialisation} :
 \STATE X$_n$ = X - min(X) / (max(X) - min(X)) %Normalize X using min-max normalization to X$_n$
 \STATE X$_i$ = round(X$_n$ $\times$ 255) %Convert X$_n$ into uint8 X$_i$
 \STATE Y$_o$ = One-Hot\_Encode(Y) %Encode Y using one-hot encoding to Y$_o$
 \STATE Initialise Weight vector W
 \STATE Binarize W
 \\ \textit{LOOP Process}
  \FOR {$i = 0$ to $N$}
    \STATE Y$_t$[B] = W $\cdot$ X$_i$
    \STATE error = Y$_t$[B] - Y$_o$[B]
    \STATE Calculate $\delta$ W using squared-hinge loss with ADAM optimizer
    \STATE Binarize W     
  \ENDFOR
 \end{algorithmic} 
 \label{algo1}
\end{algorithm}

\subsection{Weight Mapping Strategy}
\label{map1}
Proposed weight mapping strategy for using two-terminal RRAM device crossbar is shown in Fig. \ref{Fig1}. In this discussion, we primarily focus on the binarized-ADALINE as an example network since it represents a basic unit for realization of a multi-layer BNN. Every weight vector is mapped using two separate components i.e. $W^+$,$W^-$ in consecutive rows of the crossbar. If length of input vector is greater than number of columns, we partition weights and allocate them on separate set of rows. Since two devices are utilized for representing each logical weight (shown in Fig. \ref{Fig1}), the utilization of crossbar is reduced by 50\%. Summation performed at each row of the crossbar can be represented by Eq. \ref{eq1} (where $k$ denotes output class, $i$ denotes polarity i.e. +ve, -ve and $p$ denotes partition of feature vector [0,n]). Post summation stage (i.e. TIA output), final sum voltages from rows $W^+$ and $W^-$ are subtracted at the 2nd stage (Eq. \ref{eq3}) leading to a class-wise score $V_{o,k}$, which is used for generating a class decision at the final stage. In case of partitions, all output voltages of a respective polarity (i), for class k ($V^{k,i}_{o,p}$) are first stored and summed (Eq. \ref{eq2}) before proceeding to the subtraction and class-decision stages. In proposed binarized-ADALINE, we use analog inputs of 8-bit precision. 8-bit Inputs are realized by utilizing pulse-width modulation (PWM) based encoding which requires integration across 255 cycles to generate the final class score. Fig. \ref{Fig1b} summarizes sequence of operations.

\begin{align}
V_{o,p}^{k,i} = W_p^{i,k} \cdot X \label{eq1} \\    
V_{o}^{k,i} = \sum\limits_{p=1}^n V_{o,p}^{k,i}  \label{eq2} \\
V_{o,k} =  V_{o,k}^+ - V_{o,k}^- \label{eq3} 
\end{align}
\section{Experimental Results}
\label{sec4}
\begin{figure}[b]
    \centering
    \includegraphics[width=0.88\linewidth]{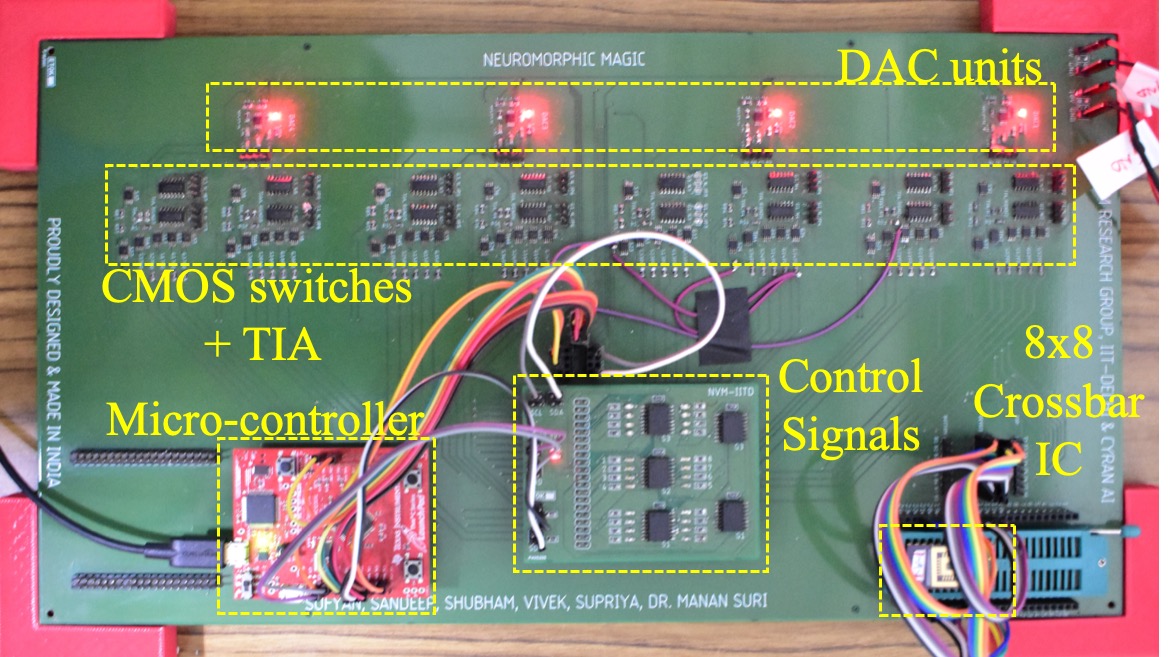}
    \caption{Custom-testbench of proposed OxRAM based VMM. Since our DAC units can only generate +ve voltage signals, -ve voltage across the RRAM device is realized by applying a +ve voltage to the bottom terminal while grounding the top terminal of the device.}
    \label{fig_setup}
\end{figure}

\begin{figure*}[ht]
\centering
  \includegraphics[width=0.95\textwidth]{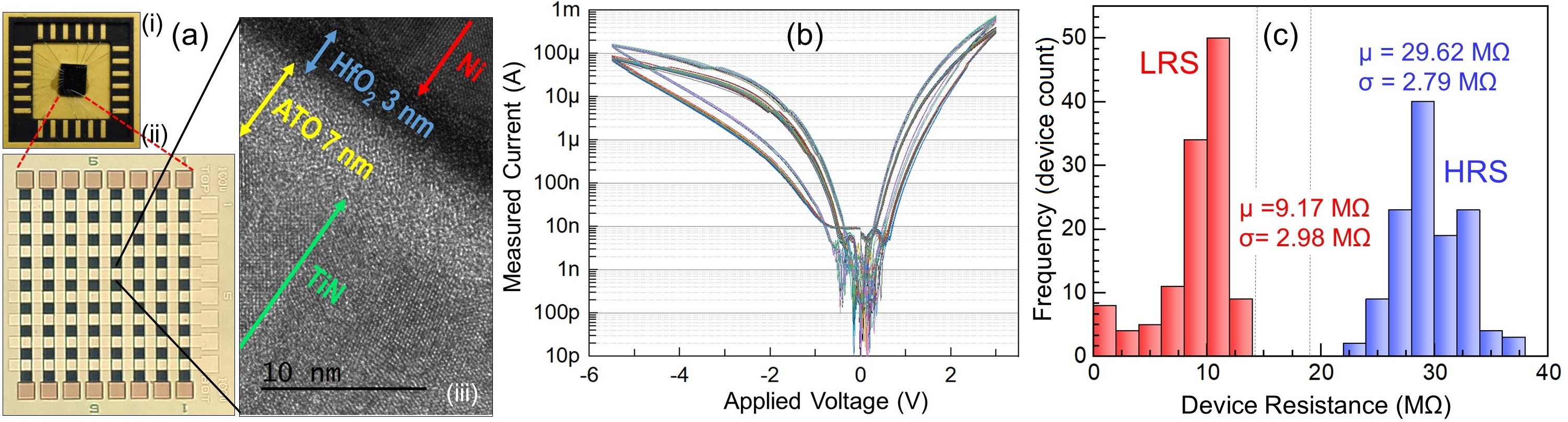}
  \caption{(a) (i) Packaged 8$\times$8 OxRAM crossbar IC used for testing. (ii) Image of the crossbar die acquired using an optical microscope. (iii) HR-TEM image of of bilayer Ni/$HfO_2$/ATO/TiN OxRAM device fabricated for this study. (b) Overlaid DC IV curves of 64 OxRAM devices in 8$\times$8 crossbar indicating low D2D variability. ($V_{set}$ = 3.3 V, $V_{reset}$ = -5.5 V) (c) Resistance distribution of LRS and HRS states used for the study (V$_{read}$ = - 0.8 V).}
  \label{Fig2}
\end{figure*}

\begin{figure*}[htbp]
\centering
\includegraphics[width=\textwidth]{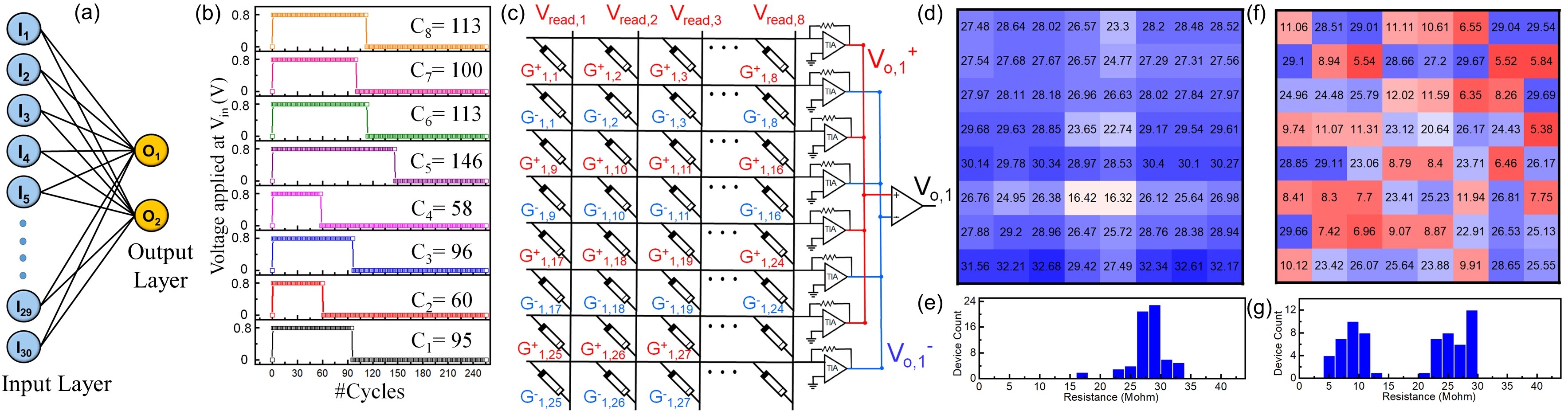}
\caption{(a) Implemented 30$\times$2 ADALINE network. (b) Proposed PWM based input encoding scheme for mapping analog inputs to crossbar columns. (c) Mapping strategy used in this study for realizing 30$\times$2 ADALINE network on a 8$\times$8 crossbar. Resistance state and distribution of 8$\times$8 OxRAM crossbar before programming (d, e) and after programming (f, g). All resistance values are in M$\Omega$.}
\label{Fig6}
\end{figure*}
\subsection{Testbench \& Dataset}
Fig. \ref{fig_setup} shows the custom-testbench built for this application. The main components are: 
\begin{enumerate}
    \item $Micro-controller$: Drives the overall VMM operation scheduling. It is also used for communication with host PC, managing control signals to all other ICs, OxRAM crossbar programming/sensing/ weight-mapping partitioning, computation of outputs post-TIA stage, final class score computation and inference decision.
    \item $Sensing\ Circuitry$: Consists of TIA for converting current to voltage signal and ADC unit (from micro-controller). 
    \item $DAC\ units$: For generating required OxRAM programming signals. We used V$_{set}$ = 3.3 V, V$_{reset}$ = -5.5 V and V$_{read}$ = -0.8 V for programming OxRAM devices in our crossbar. (High values of V$_{set}$,V$_{reset}$ are used in our prototype devices as they have large active area (100 $\times$ 100 $\mu m^2$). Negative V$_{reset}$,V$_{read}$ were realized by applying +ve voltage signals at the bottom electrode of OxRAM device and grounding the top electrode, thereby generating effective -ve V$_{TB}$ (where V$_{TB}$ = V$_{Top}$ - V$_{Bottom}$).
    \item $CMOS\ Switches$: For path selection (during SET, RESET and READ operation) and compliance current control.
\end{enumerate}
For validation of the proposed methodology, we trained the binarized-ADALINE network using a binary classification dataset (Breast Cancer dataset) from the UCI machine learning repository \cite{Dua_2019}. The dataset consists of 357 benign and 212 malignant cells.
\subsection{Fabricated OxRAM Crossbar Chip}
%We have used 8$\times$8 OxRAM crossbar. 
8$\times$8 OxRAM crossbar with resistive switching stack of Ni/3-nm HfO$_2$/7 nm Al-doped-TiO$_2$/TiN (shown in Fig. \ref{Fig2}(a)) was fabricated for this study. First, 500 nm thick TiN film was deposited on thermal-SiO$_2$ (500 nm)/Si wafer by reactive DC sputtering. The wordlines were then patterned by optical photolithography (first mask) and dry etching using inductively-coupled plasma (ICP). The bottom, 7 nm thick ATO dielectric, was then deposited by interchanging varying amount of TiO$_2$ and Al$_2$O$_3$ PE-ALD cycles. Upper, 3 nm thick dielectric HfO$_2$ film, was deposited using TDMAHf (Tetrakis(dimethylamido)hafnium) and O$_2$ plasma. All depositions were carried out at 250 $^\circ$C using remote plasma hot-wall reactor PE-ALD system. The TE pattern (similar to the BE pattern but rotated 90$^\circ$) was defined using second mask and 100 nm thick Ni top electrode film was deposited by DC sputtering and patterned  using lift-off technique. This way, an 8$\times$8 crossbar was formed with 100 $\mu$m wide perpendicular TiN and Ni wordlines and bitlines sandwiching the dielectric bilayer, forming 64 OxRAM devices with 100x100 $\mu$m$^2$ active area at each crosspoint.  The third mask and ICP dry etching step was performed to open the contact windows (etch the dielectrics) to the wordline contact pads. Wire bonding and packaging were the final steps for the OxRAM crossbar encapsulation. DC IV characteristics of 64 OxRAM devices in the 8$\times$8 crossbar are overlaid in Fig. \ref{Fig2}(b). In our study, we have used two distinguishable well separated OxRAM resistance states (HRS/LRS) as shown in Fig. \ref{Fig2}(c).  
\subsection{BNN Results on OxRAM Crossbar}
To perform BNN computation with OxRAM crossbar, row-level READ operations were used. Since we used V$_{read}$ = -0.8 V, due to negative read voltage current integration happened over the row elements. For binarized-ADALINE network (see Fig. \ref{Fig6}(a)), we partitioned and mapped the weight vector over multiple rows (shown in Fig. \ref{Fig6}(c)) and performed VMM operations. The output class was decided by Eq. \ref{eq3} as described in Section \ref{map1}. The 8-bit wide positive input values were encoded as PWM duty cycles applied to crossbar columns (Fig. \ref{Fig6}(b)) leading to current integration over time. In PWM encoding, the input value (ranging from 0 to 255) was translated to pulses with fixed voltage (0.8 V). Specific pulse-widths were n$\times$17 ms (where n= number of clock cycles [0-255]). Prior to executing VMM operations the entire crossbar was initialized to HRS (Fig. \ref{Fig6}(d,e)) for reliable weight programming. Crossbar resistance distribution post-mapping of trained weights is shown in Fig. \ref{Fig6} (f, g). Training was performed on 80\% of the UCI cancer dataset with random shuffling. Classification accuracy results are shown in Table \ref{tab1}. We can observe a reduction in classification accuracy between simulated and experimental network. The drop in accuracy can be attributed to sneak-path issue inherent to crossbars. Due to sneak-path leakage, the effective integrated current value across a specific row is diminished. Another factor contributing to accuracy loss is the variability in OxRAM device programming (i.e. LRS and HRS distributions) as evident from Fig. \ref{Fig2}(c), \ref{Fig6}(g). An effective strategy to mitigate these effects would be to implement two separate ADALINE networks in place of a single neuron. Deep binarized multi-layer networks can also be explored, using larger RRAM arrays to further improve the learning performance as shown in simulation studies \cite{hirtzlin2019implementing,Fouda_2019,Kim_2019}. It should be noted that by using binary OxRAM states (HRS/LRS), we have limited the effect of variability that would otherwise reflect in an analog resistance VMM implementation. Furthermore, using OxRAM crossbar solely for inference relaxes the endurance requirements for the devices.
\begin{table}[tb]
    \centering
    \caption{Classification accuracy for VMM-based binarized-ADALINE.}
    \label{tab1}    
    \begin{tabular}{|p{0.3cm}|p{4.4cm}|p{1.2cm}|p{1.2cm}|}
        \hline
         Sr. No. & Platform  &  Training Accuracy \% & Testing Accuracy \% \\ \hline
         1 & Simulation (Software) & 74.06 & 78.07 \\ \hline
         2 & Experimental (8$\times$8 OxRAM Crossbar) & 68.13 & 67.54 \\  \hline  
    \end{tabular}
\end{table}
\vspace{-1 pt}
\section{Conclusion}
In this paper, we presented a novel methodology for mapping BNN operations on two-terminal binary NVM device based crossbars of arbitrary sizes. We experimentally demonstrate the realization of a binarized-ADALINE classifier, for UCI Cancer dataset, on fabricated 8$\times$8 OxRAM crossbar. The demonstrated methodology can be extended to deeper BNNs through proposed energy-efficient in-memory VMM computation strategy.  
\vspace{-1 pt}
\section*{Acknowledgement}
This work was supported in part by SERB-CRG/2018/001901 and CYRAN AI Solutions.
\bibliographystyle{IEEEtran}
\bibliography{ref}
\end{document}